\documentclass[conference]{IEEEtran}

\usepackage[utf8]{inputenc}
\PassOptionsToPackage{table}{xcolor}
\usepackage{amsmath}
\usepackage{listings}
\usepackage{amssymb}
\usepackage{amsmath}
\usepackage{graphicx}
\usepackage{makeidx}
\usepackage{multicol}
\usepackage{multirow}
\usepackage{makecell}
\usepackage{mathptmx}
\usepackage{xcolor}
\usepackage{xspace}
\usepackage{booktabs}
\usepackage{acronym}
\usepackage{todonotes}
\usepackage{url}
\usepackage{breakurl}
\usepackage[breaklinks]{hyperref}
\usepackage{comment}
\usepackage{longtable}
\usepackage{tikz}
\usepackage{mathdots}
\usepackage{yhmath}
\usepackage{cancel}
\usepackage{color}
\usepackage{siunitx}
\usepackage{array}
\usepackage{multirow}
\usepackage{amssymb}
\usepackage{gensymb}
\usepackage{tabularx}
\usepackage{xcolor}
\usetikzlibrary{fadings}
\usepackage[noend]{algpseudocode}
\usepackage{lscape} 
\usepackage{color, colortbl}
\usepackage[sc]{mathpazo} 
\usepackage[T1]{fontenc} 
\usepackage{microtype} 
\usepackage{multicol} 
\usepackage{booktabs} 
\usepackage{float} 
\usepackage{hyperref} 
\usepackage{lettrine} 
\usepackage{paralist} 
\usepackage{todonotes}
\usepackage{url}
\usepackage[export]{adjustbox}
\usepackage{tcolorbox}
\usepackage{algpseudocode}
\usepackage{algorithm}
\usepackage{tikz}
\usepackage{multirow}
\usepackage{cancel}
\usepackage{color}
\usepackage{siunitx}
\usepackage{array}
\usepackage{multirow}
\usepackage{amssymb}
\usepackage{gensymb}
\usepackage{tabularx}
\usepackage{xcolor}

\usepackage{breqn}

\usetikzlibrary{shadows,trees}

\definecolor{lightgray}{rgb}{.9,.9,.9}
\definecolor{mauve}{rgb}{0.58,0,0.82}
\definecolor{darkgray}{rgb}{.4,.4,.4}
\definecolor{darkgreen}{rgb}{0, 0.39, 0.00}
\definecolor{brinkpink}{rgb}{0.98, 0.38, 0.5}
\definecolor{brightpink}{rgb}{1.0, 0.0, 0.5}
\definecolor{Gray}{gray}{0.7}

\acrodef{CIA}{Change Impact Analysis}
\acrodef{OSS}{Open Source Software}
\acrodef{RCC}{Rapid Release Change}
\acrodef{WALA}{T. J. Watson Libraries for Analysis}
\acrodef{ARA}{Architectural Risk Analysis} 
\acrodef{SecGraph}{security graph}
\acrodef{PCI}{Payment Card Industry}
\acrodef{NIST}{The National Institute of Standards and Technology}
\acrodef{OT}{Operational Technology}
\acrodef{IT}{Information Technology}
\acrodef{CPS}{Cyber-physical System}
\acrodef{OEM}{Original Equipment Manufacturer}
\acrodef{SWOT}{Strength, Weaknesses, Opportunity, and Threats}

\begin{document}

\title{Threat Modeling of Cyber-Physical Systems in Practice} 

\author{\IEEEauthorblockN{ Ameerah-Muhsinah Jamil\IEEEauthorrefmark{1}~Lotfi ben Othmane
\IEEEauthorrefmark{1}Altaz Valani\IEEEauthorrefmark{2}}

\\
\IEEEauthorblockA{\IEEEauthorrefmark{1} Iowa State University, Ames, IA USA~\IEEEauthorrefmark{2}Security Compass, Canada
}}

\maketitle

\begin{abstract}

Traditional \acp{CPS} were not built with cybersecurity in mind. They operated on separate \ac{OT} networks. As these systems now become more integrated with \ac{IT} networks based on IP, they expose vulnerabilities that can be exploited by the attackers through these IT networks. The attackers can control such systems and cause behavior that jeopardizes the performance and safety measures that were originally designed into the system. In this paper, we explore the approaches to identify threats to \acp{CPS} and ensure the quality of the created threat models. The study involves interviews with eleven security experts working in security consultation companies, software engineering companies, an \ac{OEM}, and ground and areal vehicles integrators. We found through these interviews that the practitioners use a combination of various threat modeling methods, approaches, and standards together when they perform threat modeling of given \acp{CPS}. Key challenges practitioners face are: they cannot transfer the threat modeling knowledge that they acquire in a cyber-physical domain to other domains, threat models of modified systems are often not updated, and the reliance on mostly peer-evaluation and quality checklists to ensure the quality of threat models. The study warns about the difficulty to develop secure \acp{CPS} and calls for research on developing practical threat modeling methods for \acp{CPS}, techniques for continuous threat modeling, and techniques to ensure the quality of threat models.

\end{abstract}

\section{Introduction}

In the past, \acp{CPS} operated on their own networks, which were separated or air-gapped from the corporate IT networks. The OT and IT networks started converging in response to the need to provide data and insights to stakeholders on IT networks. The challenge with integrating these technologies is the velocity of change: IT technologies tend to change very frequently, and updates or patches can be readily done while OT technologies have a considerably longer shelf life. Legacy security concerns when OT technologies were initially deployed can be significantly different from the present security concerns. Trying to capture this disparity is done, in part, through threat modeling. 

Until recently, attackers needed physical access to \acp{CPS}. The trend of integrating these systems to IP networks and the internet for services, such as remote car diagnostic and cooperative adaptive cruise control, has extended the attack surface. The goals of attacks on cyber-physical systems, such as Stuxnet and Triton, are often not to breach the confidentiality, integrity, or availability of the system's data but to make the target system perform activities other than the ones planned and expected by the original designers. Hence changing the actual process and unleaching damaging consequences.

Threat modeling is a "systematic exploration technique to expose any circumstance or event having the potential to cause harm to a system in the form of destruction, disclosure, modification of data, or denial of service"~\cite{ieee}. It is an approach for identifying potential threats to a given software and suggesting mitigations. In this paper, we will not discuss mitigations and limit the scope to threats only.

There are several methods for threat modeling, including threat tree, attack tree, STRIDE, and abuse cases~\cite{Sho2014}. Xiong and Lagerstrom surveyed threat modeling literature. The authors of many of the surveyed papers validated their proposed approaches (22 out of the 54 selected papers) using, for example, case studies, and simulation while only two papers used real-word applications~\cite{XIONG201953}. Most of these methods have been designed for information systems where the assets are data at rest and in-transit. The focus on data within the IT network is an important one. Threat modeling of \ac{OT} components can often be physically dangerous, expensive, or even unrealistic. Therefore, looking at the problem from a data-centric perspective can help to identify the data flow early before it gets to the OT technology but may not be sufficient to identify misuses of \acp{CPS}.

There is a lack of systematic threat modeling methods for \acp{CPS} and a gap between the literature and the practice of threat modeling of \acp{CPS}. This paper aims to address that by answering the question: \emph{What are the practices of threat modeling of cyber-physical systems by cyber-security experts?} To address this question, we interviewed eleven security experts who perform threat modeling of cyber-physical systems in their respective organizations. Then, we transcribed the interviews, extracted the main information, and grouped them into themes, and analyzed the findings. We found that:
\begin{enumerate}
\item the
practitioners use a combination of threat modeling methods, approaches, and standards, together, when performing threat modeling of cyber-physical systems,

\item organizations often do not update the threat models of their modified cyber-physical systems, 

\item there is no effective method for ensuring the quality of threat models besides peer-evaluation and quality checklists.

\item the practitioners face several challenges when performing threat modeling of cyber-physical systems, including the difficulty to transfer the threat modeling knowledge they acquire in a cyber-physical domain to other domains.
\end{enumerate}

The results of this work could be used by organizations when performing threat modeling of cyber-physical systems and by academia to develop solutions and techniques that help practitioners perform threat modeling efficiently.

This paper is organized as follows: Section~\ref{sec:relwork} discusses related works, Section~\ref{sec:datacollection} describes the research approach, Section~\ref{sec:datananalysis} presents the results of the study, Section~\ref{sec:Discussion} summarizes the study results and discusses the impacts and limitations of the study, and Section~\ref{sec:Conclusion} concludes the paper.

\section{Related Work}\label{sec:relwork}
This section discusses related work on the security of \acp{CPS} and threat modeling methods and standards.

\noindent {\bf Security of \acp{CPS}.} Security issues of \acp{CPS} has been studied for several years. For instance, Alguliyev et al.~\cite{ALGULIYEV2018212} analyzed the main types of attacks and threats of \acp{CPS} and proposed a tree of attacks that includes the attacks on sensing, actuation, computing, communication, and feedback loops; Lu et al.~\cite{LuTianbo} proposed a framework of \acp{CPS} security, which includes the security objectives, approaches, and applications of \acp{CPS}; and Pakizeh proposed a framework that aims to understand the cyber attacks and related risk of different elements of \acp{CPS}~\cite{morteza}. In addition, using the expert knowledge on security aspects, such as the form of attacks, attacker positions, operating systems, and routing permissions klaudel and Rataj~\cite{klaudel} proposed an attack graph that describes the software and hardware of a \ac{CPS} and their mutual mapping with security artifacts and a workflow that automates the construction of a vulnerability model of a \ac{CPS} that is used to quantitatively analyze the threat models of the \acp{CPS}, and estimate their exploitation costs.

The concern in security in \ac{IT} is the reduction of monetary losses and is the safety of people and controllability of the systems, besides the reduction of monetary losses, in the case of \acp{CPS}~\cite{6728305}. Sabaliauskaite and Mathur~\cite{giedre} proposed the integration of safety and security life-cycle processes and a model that unifies the attack tree and the fault tree and their countermeasures. Dong et al.~\cite{dong} proposed security and safety framework, and security framework that focuses on the security of information and controllability of the \acp{CPS}.

\ac{NIST} developed a \ac{CPS} framework to assist in developing secure and safe \acp{CPS}~\cite{nist}. The security concern of the framework is to protect \acp{CPS} from unauthorized accesses, change damages, and destruction in addition to the CIA triad, and the safety concern is preventing negative consequences of cyber attacks on the stakeholders, including life, health, property, data, and damage to the physical environment.

\noindent{\bf Threat Modeling methods and standards.} There exist several works on threat modeling for \acp{CPS}~\cite{XIONG201953}. For instance, Martins et al.~\cite{Martins2015TowardsAS} proposed a tool for systematic analysis of threat models that includes sketching metamodel of the system using GME, defining the data-flow and its attribute, and identifying the vulnerabilities that may exist in the data-flow connections. Also, Khan et al.~\cite{khan} adapted the STRIDE method for \acp{CPS} by focusing on the data-flow between the components of the system, which demonstrated promising results when applied to a case study as it identifies the vulnerabilities at cyber sub-systems and their potential consequences on the physical components of the system. In addition, Casola et al.~\cite{CASOLA2019100056} developed a threat catalog that consists of known threats affecting different components of IoT and classified them based on asset types.

Several researchers acknowledged the impact of application domains on the threat modeling of \acp{CPS}. For instance, Meyer et al.~\cite{meyer} proposed an attack tree to threat model building and home automation systems in order to identify security faults either in implementation or deployment, and Suleiman et al.~\cite{SULEIMAN2015147} developed a comprehensive threat modeling by integrating the results of smart grid system security threat analysis with the reference architecture of smart grid including the components and communication among them.

The International Standards Organization (ISO) and SAE International released standard ISO/SAE 21434- Road vehicles cybersecurity engineering to address the need in cybersecurity engineering of electrical and electronic systems within road vehicles. The standard provides guidelines to integrate cybersecurity concerns in product development, and perform cybersecurity assessment and monitoring, and develop policies to handle cybersecurity incidents.

This paper addresses the gap between the development of threat modeling methods, techniques, and standards and the practice of threat modeling of \acp{CPS}. 

\section{Research Approach}\label{sec:datacollection}
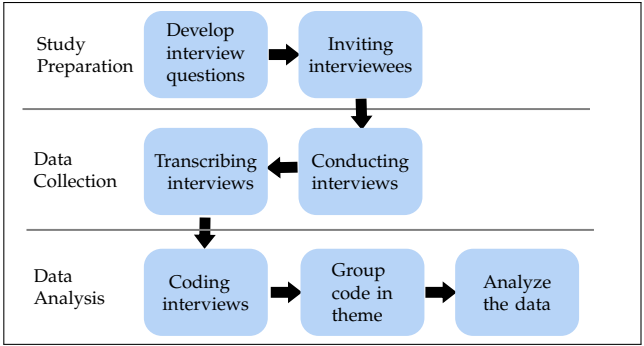
\begin{figure}
    \centering
    \fbox{
    \begin{minipage}{8cm}

\tikzset{every picture/.style={line width=0.75pt}} 

\begin{tikzpicture}[x=0.75pt,y=0.75pt,yscale=-1,xscale=1]

\draw  [color={rgb, 255:red, 183; green, 212; blue, 247 }  ,draw opacity=1 ][fill={rgb, 255:red, 183; green, 212; blue, 247 }  ,fill opacity=1 ] (81,27.8) .. controls (81,23.05) and (84.85,19.2) .. (89.6,19.2) -- (133.9,19.2) .. controls (138.65,19.2) and (142.5,23.05) .. (142.5,27.8) -- (142.5,53.6) .. controls (142.5,58.35) and (138.65,62.2) .. (133.9,62.2) -- (89.6,62.2) .. controls (84.85,62.2) and (81,58.35) .. (81,53.6) -- cycle ;
\draw [color={rgb, 255:red, 128; green, 128; blue, 128 }  ,draw opacity=1 ]   (19.3,68.2) -- (305.5,68.2) ;

\draw  [color={rgb, 255:red, 0; green, 0; blue, 0 }  ,draw opacity=1 ][fill={rgb, 255:red, 0; green, 0; blue, 0 }  ,fill opacity=1 ][line width=0.75]  (144,38.45) -- (152.7,38.45) -- (152.7,36.2) -- (158.5,40.7) -- (152.7,45.2) -- (152.7,42.95) -- (144,42.95) -- cycle ;
\draw  [color={rgb, 255:red, 0; green, 0; blue, 0 }  ,draw opacity=1 ][fill={rgb, 255:red, 0; green, 0; blue, 0 }  ,fill opacity=1 ][line width=0.75]  (157.55,99.8) -- (148.85,99.98) -- (148.89,102.23) -- (143,97.85) -- (148.71,93.23) -- (148.75,95.48) -- (157.45,95.3) -- cycle ;
\draw  [color={rgb, 255:red, 0; green, 0; blue, 0 }  ,draw opacity=1 ][fill={rgb, 255:red, 0; green, 0; blue, 0 }  ,fill opacity=1 ][line width=0.75]  (223,158.45) -- (231.7,158.45) -- (231.7,156.2) -- (237.5,160.7) -- (231.7,165.2) -- (231.7,162.95) -- (223,162.95) -- cycle ;
\draw  [color={rgb, 255:red, 0; green, 0; blue, 0 }  ,draw opacity=1 ][fill={rgb, 255:red, 0; green, 0; blue, 0 }  ,fill opacity=1 ][line width=0.75]  (113.61,123.49) -- (113.48,132.19) -- (115.73,132.22) -- (111.13,137.95) -- (106.73,132.08) -- (108.98,132.11) -- (109.12,123.42) -- cycle ;
\draw  [color={rgb, 255:red, 0; green, 0; blue, 0 }  ,draw opacity=1 ][fill={rgb, 255:red, 0; green, 0; blue, 0 }  ,fill opacity=1 ][line width=0.75]  (192.3,63.39) -- (192.54,72.09) -- (194.79,72.03) -- (190.45,77.95) -- (185.79,72.27) -- (188.04,72.21) -- (187.8,63.51) -- cycle ;
\draw  [color={rgb, 255:red, 0; green, 0; blue, 0 }  ,draw opacity=1 ][fill={rgb, 255:red, 0; green, 0; blue, 0 }  ,fill opacity=1 ][line width=0.75]  (145,158.45) -- (153.7,158.45) -- (153.7,156.2) -- (159.5,160.7) -- (153.7,165.2) -- (153.7,162.95) -- (145,162.95) -- cycle ;
\draw  [color={rgb, 255:red, 183; green, 212; blue, 247 }  ,draw opacity=1 ][fill={rgb, 255:red, 183; green, 212; blue, 247 }  ,fill opacity=1 ] (159,27.8) .. controls (159,23.05) and (162.85,19.2) .. (167.6,19.2) -- (211.9,19.2) .. controls (216.65,19.2) and (220.5,23.05) .. (220.5,27.8) -- (220.5,53.6) .. controls (220.5,58.35) and (216.65,62.2) .. (211.9,62.2) -- (167.6,62.2) .. controls (162.85,62.2) and (159,58.35) .. (159,53.6) -- cycle ;
\draw  [color={rgb, 255:red, 183; green, 212; blue, 247 }  ,draw opacity=1 ][fill={rgb, 255:red, 183; green, 212; blue, 247 }  ,fill opacity=1 ] (81,86.8) .. controls (81,82.05) and (84.85,78.2) .. (89.6,78.2) -- (133.9,78.2) .. controls (138.65,78.2) and (142.5,82.05) .. (142.5,86.8) -- (142.5,112.6) .. controls (142.5,117.35) and (138.65,121.2) .. (133.9,121.2) -- (89.6,121.2) .. controls (84.85,121.2) and (81,117.35) .. (81,112.6) -- cycle ;
\draw  [color={rgb, 255:red, 183; green, 212; blue, 247 }  ,draw opacity=1 ][fill={rgb, 255:red, 183; green, 212; blue, 247 }  ,fill opacity=1 ] (81,147.8) .. controls (81,143.05) and (84.85,139.2) .. (89.6,139.2) -- (133.9,139.2) .. controls (138.65,139.2) and (142.5,143.05) .. (142.5,147.8) -- (142.5,173.6) .. controls (142.5,178.35) and (138.65,182.2) .. (133.9,182.2) -- (89.6,182.2) .. controls (84.85,182.2) and (81,178.35) .. (81,173.6) -- cycle ;
\draw  [color={rgb, 255:red, 183; green, 212; blue, 247 }  ,draw opacity=1 ][fill={rgb, 255:red, 183; green, 212; blue, 247 }  ,fill opacity=1 ] (159,86.8) .. controls (159,82.05) and (162.85,78.2) .. (167.6,78.2) -- (211.9,78.2) .. controls (216.65,78.2) and (220.5,82.05) .. (220.5,86.8) -- (220.5,112.6) .. controls (220.5,117.35) and (216.65,121.2) .. (211.9,121.2) -- (167.6,121.2) .. controls (162.85,121.2) and (159,117.35) .. (159,112.6) -- cycle ;
\draw  [color={rgb, 255:red, 183; green, 212; blue, 247 }  ,draw opacity=1 ][fill={rgb, 255:red, 183; green, 212; blue, 247 }  ,fill opacity=1 ] (160,147.8) .. controls (160,143.05) and (163.85,139.2) .. (168.6,139.2) -- (212.9,139.2) .. controls (217.65,139.2) and (221.5,143.05) .. (221.5,147.8) -- (221.5,173.6) .. controls (221.5,178.35) and (217.65,182.2) .. (212.9,182.2) -- (168.6,182.2) .. controls (163.85,182.2) and (160,178.35) .. (160,173.6) -- cycle ;
\draw  [color={rgb, 255:red, 183; green, 212; blue, 247 }  ,draw opacity=1 ][fill={rgb, 255:red, 183; green, 212; blue, 247 }  ,fill opacity=1 ] (238,147.8) .. controls (238,143.05) and (241.85,139.2) .. (246.6,139.2) -- (290.9,139.2) .. controls (295.65,139.2) and (299.5,143.05) .. (299.5,147.8) -- (299.5,173.6) .. controls (299.5,178.35) and (295.65,182.2) .. (290.9,182.2) -- (246.6,182.2) .. controls (241.85,182.2) and (238,178.35) .. (238,173.6) -- cycle ;
\draw [color={rgb, 255:red, 128; green, 128; blue, 128 }  ,draw opacity=1 ]   (21.3,129.2) -- (307.5,129.2) ;

\draw (111.75,40.7) node [scale=0.7] [align=left] { Develop \\ interview \\ questions};
\draw (111.75,99.7) node [scale=0.7] [align=left] { Transcribing \\ \ \ interviews};
\draw (189.75,40.7) node [scale=0.7] [align=left] { \ \ \ Inviting \\interviewees};
\draw (111.75,160.7) node [scale=0.7] [align=left] { \ Coding\\interviews};
\draw (268.75,160.7) node [scale=0.7] [align=left] {Analyze\\the data};
\draw (190.75,160.7) node [scale=0.7] [align=left] { Group \\code in \\ theme};
\draw (189.75,99.7) node [scale=0.7] [align=left] {Conducting\\ interviews};
\draw (51,42) node [scale=0.7] [align=left] {Study\\Preparation};
\draw (46,99) node [scale=0.7] [align=left] {Data\\Collection};
\draw (43,159) node [scale=0.7] [align=left] {Data \\Analysis};

\end{tikzpicture}
\end{minipage}
}
\caption{Phases of the study.} 
\label{fig:researchApproach}
\vspace{-.15in}
\end{figure}

This study aims to explore the practice of threat modeling of \acp{CPS} in the industry. The data source of the study comes from interviewing a set of security experts practicing threat modeling. Figure \ref{fig:researchApproach} illustrates the process of the study, which has three phases: study preparation, data collection, and data analysis. The descriptions of the phases follow.

\subsection{Study Preparation}
The description of the study preparation follows.

\noindent{\bf Interview protocol.} We reviewed the literature on threat modeling of a \ac{CPS}. We used the knowledge that we acquired to develop a questionnaire protocol. We specified the research goal with the project sponsor and formulated a set of open-ended interview questions. The questionnaire was tested by trial runs with team members and revised based on the feedback. The set of questions consists of eleven open-ended questions--Open-ended questions encourage the participants to provide detailed responses.

\noindent{\bf Participants selection.} We invited a set of security experts working in cyber-security companies. Eleven participants accepted our requests and participated in the study with the goal to contribute to science, not to represent their employers. Table \ref{tab:business} shows the experience of each participant on threat modeling and the business of their employers. Among the participants, three work for major software development companies and five work for major companies that develop \acp{CPS}.    

\begin{table}
\centering
\caption{Business of each participant.}\label{tab:business}
\rowcolors{2}{gray!25}{white}
\renewcommand{\cellalign}{l}
\small
\begin{tabular}{p{0.6in} p{2.2in}}
\hline 
\rowcolor{gray!50}
Participant &  Business \\
\hline
P1 &Security consultation \\
P2  &Software engineering \\
P3  &Security consultation \\
P4  &Areal vehicles integrator \\
P5  &Software engineering \\
P6  &Software engineering \\
P7  &Security consultation \\
P8  & Ground vehicles integrator \\
P9  & Ground vehicles integrator \\
P10 & Original Equipment Manufacturer (OEM) \\
P11  & Ground vehicles integrator \\
\hline 
\end{tabular}
\vspace{-4mm}
\end{table}

\begin{table}
\centering
 \caption{Threat modeling themes.}\label{tab:codes}
\rowcolors{2}{gray!25}{white}
\small
\begin{tabular}{p{0.7in} p{2.0in} }
\hline
\rowcolor{gray!50}

Theme & Description \\
 \hline 
Security aspects & It concerns confidentiality, integrity, and availability.  \\

Threat business impacts & The other aspects that the participant is concerned about when performing threat modeling including users' safety and company reputation.\\

Threat modeling approaches & The approaches and methods that the participants use for threat modeling, e.g., asset-centric, attacker-centric, STRIDE etc. \\

Threat identification methods & The methods that the participants use to identify the threats which is part of the threat modeling process.\\

Threat modeling steps & The activities or steps performed by the experts to identify the threat model of a given system.\\

Continuous Threat modeling & The process used to update threat models to address system changes. \\

Quality assurance of threat models & The methods used to assess and evaluate the quality of the threat models. \\

Tools & The tools used in the threat modeling process. \\

Involved people & People involved in the threat modeling process.\\

Challenge & The challenge that experts face when performing threat modeling for cyber physical systems.\\

Suggestion & Suggestions to improve the threat modeling process for cyber physical systems.\\
 
 \hline 
\end{tabular}
\end{table}

\subsection{Data Collection}
The data collection consists of two sub-phases: conducting the interviews and transcribing the interviews. The descriptions of these sub-phases follow.

\noindent{\bf Conducting the interview.} We scheduled a one-hour meeting with each expert. The meetings were held through Zoom and Web-ex because the interviewers and participants are located in different places. The interviews were conducted by one of the authors. The interviewer explained to each of the interviewees at the beginning of each of the meetings the goal of the project, the interview process and requested the consent of the participant to record the interview.

\noindent{\bf Transcription of the interviews.} The interviews were transcribed using oTranscribe \footnote{oTranscribe: \url{https://otranscribe.com/}} and Otter.ai. \footnote{Otter.ai: \url{https://otter.ai/}}.

\subsection{Data Analysis}

\noindent{\bf Interview coding.} We used the thematic analysis method for the interview coding \cite{saldana2012coding}. Thematic analysis is "a method for identifying, analyzing and reporting patterns within data"~\cite{BrCl2006}. It allows researchers to explore phenomena through interviews, stories, and observations~\cite{Con2010}.

Interview coding uses the interview transcripts as the input and outputs codes that identify the aspects mentioned during the interviews. A code is a word or short phrase identifying the essence of a portion of text. At the end of this step, we assigned codes to each of the eleven interview transcripts. For example, we assigned code \textit{security properties/goal} to the text \textit{"When it comes to the \acp{CPS}, the availability of the system matters a lot"}. Codes that were semantically similar across transcripts were consolidated. We used Atlas.ti \footnote{ATLAS.ti: \url{https://atlasti.com/}} tool to code the interviews.

\noindent{\bf Data extraction and classification.} Similar codes are grouped into themes. A theme generalizes a set of codes belonging to a given concept. The process of assigning themes to codes was done for each transcript. Table \ref{tab:codes} lists the themes and associated categories.

\noindent{\bf Analysis of the results.} From the code groups, we identified information on security properties, threat business impacts, threat modeling approaches, and method, threat modeling details activity, continuous threat modeling approach, threat identification methods, continuous threat modeling approaches, risk assessment approaches, quality assurance approaches, roles involved in threat modeling, tools, and challenges. We then modeled the relationships among these themes.

\section{Data analysis}\label{sec:datananalysis}
This section describes the themes that we extracted from the eleven interviews.

We used \textit{Pi} to refer to participant \textit{i} in the interview.

\subsection{Security Properties}

Security experts focus on protecting the confidentiality, integrity, and availability (CIA triad) of the data managed by their systems. Table \ref{tab:properties} lists the number of participants that discussed each of the security properties. We observe that the participants are concerned about data integrity and availability but not about data confidentiality. They are also concerned about secure modification, availability, consistency, accuracy, and misuses of the data over their life-cycle in their system. For instance, P9 said: \textit{"so things that are important to us are maybe not, as you said, the confidentiality of it if you're talking about a control system, but you're looking at the integrity of the messaging[...] you know, the data is the control message."} The reason is: data is used to process the control commands of the physical components of \acp{CPS}. Modification and misuses of these data can cause damages or losses, and unavailability of data and system components could prevent real-time feedback behaviors of certain \acp{CPS} and cause losses and damages. 

\begin{table}
\centering
\caption{No. of participants concerned with each of the security properties/goal. }\label{tab:properties}
 \small
\rowcolors{2}{gray!25}{white}
\renewcommand{\cellalign}{l}
\begin{tabular}{p{2.0in} p{0.9in}}
\hline
\rowcolor{gray!50}
Security properties & \# Participants\\
\hline 

Confidentiality  &  1\\
Integrity & 6 \\
Availability & 6 \\
\hline 
\end{tabular}
\vspace{-4mm}
\end{table}

\subsection{Threat Business Impacts}

Many \acp{CPS}, including connected cars, involve human as users and are safety-critical systems. Security and safety are closely related in these systems \cite{sabaliauskaite}. The exploitation of systems' weaknesses and vulnerabilities could have a high impact on the safety of the users.
For example, P3 said: \textit{"..the cyber threats can actually impact the physical safety of workers, or you know, cause an explosion within a plant or any number of potential outcomes"}. Besides safety, financial losses, and reputation damage are also important aspects that participants consider when performing threat modeling of \acp{CPS}. Security weaknesses in the supply chain is a typical example.

\subsection{Threat Modeling Approaches}

The participants in the study have either control systems or IT background. The participants with control systems background focus on the malicious controllability of the physical components of the studied system as P11 said \emph{"All these methodologies started from this classic [Referring to ISO27005] as an approach with slight modifications. What was added by Evita is the notion of controllability"}. \textbf{P1}, for example, uses a field-tested custom engine derived from the ISA/IEC 62443 standard~\cite{isa62443} to identify the physical/cyber threats that apply to each of the assets, zones (a group of assets), and conduits of the system under consideration, keeping in mind that a cyber threat can have a physical attack surface ,and \textbf{P2} uses the STRIDE taxonomy~\cite{khan} and analyze the failure scenarios that might apply to the components considering the behavior of the physical components and the safety of the system. In general, these participants combine the use of the known approaches such as STRIDE or PASTA with the analysis of failure modes and criticality of the physical systems. 

Participants with IT background apply the classic threat modeling approaches such as STRIDE~\cite{khan} and DREAD~\cite{MaCo2018}. They identify the assets, the components, and the data managed by the studied system and focus mostly on threats to the integrity, availability, and confidentiality of the data. For example, \textbf{P5} approach is: understand the system, identify the weaknesses, identify potential attacks and mitigations, and prioritize the identified threats. They consider that each \ac{CPS} operates in a specific environment, is associated with specific weaknesses and type of attacks, which justifies the use of threats on data rather than misbehavior of the components of the studied system.

Most of the participants decompose the system being analyzed into components and analyze the threats to each of the components. Participant \textbf{P7} deviates from this approach and analyze the studied system as a whole.\footnote{This approach is similar to the approach used to improve business processes~\cite{Gol2004}.} They look at the weaknesses related to the integration of the components of the given system.  

\begin{table}
\centering
\caption{No. of participants used known methods for threat modeling. }\label{tab:method}
\rowcolors{2}{gray!25}{white}
\small
\renewcommand{\cellalign}{l}
\begin{tabular}{p{1.2in} p{.5in} p{1.0in}}
\hline 

\rowcolor{gray!50}
Method & Ref. &\# Participants\\
\hline 

Attack tree & \cite{attack-tree}&  1\\
DREAD & \cite{MaCo2018}&  1 \\
EVITA or variant of&\cite{evita20119}&2\\
LINDDUN & \cite{linddun}&  1 \\
PASTA & \cite{pasta}&  1 \\
STRIDE  &\cite{khan}&  6\\
\hline 
\end{tabular}
\vspace{-4mm}
\end{table}

\subsection{Threat Identification Methods}

Threat identification, a key process in threat modeling, allows identifying the weaknesses of a given system that could cause harm and damage when exploited by attackers. Table~\ref{tab:method} provides the frequency of using the common individual threat modeling methods by the participants. The participants use (1) Known methods, such as attack-tree and STRIDE, (2) a combination of known methods, and (3) a combination of security standards and known approaches. 

\noindent{\bf Known method.} Several participants reported that they use known methods such STRIDE, PASTA~\cite{pasta}, LINDDUN~\cite{linddun}, and attack-tree~\cite{attack-tree}. Most of the participants (6 out of 11) use STRIDE. One expert mentioned that they use the attack-tree method because of its ability to cover all entry points of the attacks. Hence, they can identify all possible threats to the system. Some participants start with a known method and then elaborate further on their threat model based on their experience and knowledge. For example, participant \textbf{P2} identify the data flow diagram and the physical locations of the components of the studied system and apply the STRIDE method to identify the initial list of threats.

\noindent{\bf Combination of known methods and approaches.} Some participants reported the use of multiple approaches, such as asset-centric and attacker-centric, in the same project because they believe that each of the approaches and methods gives a different perspective of the system weaknesses and using a set of methods, although time-consuming, helps to identify the "complete" list of threats to a given system. 

\noindent{\bf Combination of threat modeling standards and known approaches.} One Participant, \textbf{P1}, uses real-world experience jointly with the ISA/IEC 62443 standard\cite{isa62443}to identify the physical/cyber threats that apply to each of the assets or zones (a group of assets).

\subsection{Continuous Threat Modeling Approaches}

Developers often modify parts of their \acp{CPS}~\cite{predicting} to introduce new features, fix existing defects, or improve the maintainability and the performance of these systems. The evolution of a system often involves changes to its components, which could invalidate the initial threat model since the changes could modify the attack surface and introduce new threats to the system. 

Some participants do not have processes and/or experience with managing the evolution of the threat models of their systems. For instance, one participant reported that they do not need to have processes for revising threat models as they are not involved in the businesses of the systems that they perform threat modeling of and another participant reported that they do not review the threat models of their systems even if these systems change. In addition, Participant {\bf P11} reported that the manufacturers of cars cannot do a correct continuous threat modeling. They said \emph{"..you have two updates per year for the cars...the information flow concerning various threats is not so good today because car manufacturers are not aware about all the threats related to the parts coming from their suppliers."} 

The rest of the participants (eight from eleven) have processes or approaches to manage continuous threat modeling. For example, Participant \textbf{P1} identifies the changes or triggers to a system under consideration and does a thorough threat and vulnerability assessment update, re-assessing the attack surfaces/sources and the related impacts, and adding new threats and vulnerabilities if necessary; Participant \textbf{P5} performs threat modeling as an activity of their adapted scrum~\cite{TaIk1986} process; Participant \textbf{P6} uses version control on source code of the software to identify changes and periodically assess in collaboration with the architect the the potential impacts of the changes on the threat model of the given system; Participant \textbf{P7} performs a full threat modeling of new systems and partial threat modeling when new components are added to existing systems (only the new components and impacted components are considered the partial threat modeling); Participant \textbf{P8} assesses the exploitability of the threats of changed systems and updates the priority of addressing the threats accordingly; and Participant \textbf{P9} uses a questionnaire to assess the impacts of the software changes on the previous ranking of the threats to the their system. We note that some participants report that they perform continuous threat modeling only for formality: to pass their systems to the next phase of the DevOps~\cite{7784617}.

We observe that most of the participants practice continuous threat modeling, and there is no common continuous threat modeling approach. This mixed input shows the importance of continuous threat modeling of \ac{CPS} for the industry and the lack of rigorous and efficient approaches to do so.

\subsection{Risk Assessment Approaches}

The participants reported the use of several risk analysis and scoring approaches, which we discuss in the following.

\noindent{\bf Using risk standard and/or regulations} P1 uses risk assessment standards ISA/IEC 62443~\cite{isa62443}, which provides guidelines to organize and facilitate a cyber security risk assessment for industrial automation
and control systems (IACS) while considering the necessary regulations and sector's security/risk specifics, and Participant P7 considers the impacts of the threats on the compliance with the regulations that their products must adherent to. For instance, P7 said \textit{Regulations play a major role in telling [..] the stakeholders what's more important to sustain the [business], right. I mean, basically, the products [could] fail [because of] the regulator, and you could be out of the business."} 
 
\noindent{\bf Known approach.} Many of the participants use common risk assessment approaches, such as FAIR~\cite{fair} and Bug Bar~\cite{simone2020}. The bug bar method, for example, requires assessing the criticality and severity of the threats in collaboration with the customer (which allows considering their concerns) and prioritize the threats based on their severity levels. The FAIR method allows using FAIR data to analyze and highlight the threats of the threat model. For instance, Participant P9 said \textit{" So we use the fair [...] threat modeling to highlight the threats and then run that in fair to actually turn that into a risk."}

\noindent{\bf In-house risk assessment methods.} Three participants have their own risk assessment methods. For instance, Participant {\bf P2} uses a risk register to report the risks of a given system and continuously monitor these risks, and Participants {\bf P8} uses a custom formula to compute the risks of a system using the revenue generated by the system and the criticality of the threats. 

\subsection{Quality Assurance Approaches}

Most of the participants reported that the quality of threat modeling exercises depends on the experience and skills of the experts who perform the threat modeling and the thoroughness of the assessment, including the detailed level of the used architecture and profoundness of the interviews with the stakeholders of the given system. For instance, {\bf P1} said \textit{"the ISA/IEC 62443 standard provides the basic framework but most of the quality of the assessments is based on real-world experience, which also helps with the quality of the specific deviations for every different sector"}

and {\bf P11} said "the expert, nothing else." Few participants use techniques to ensure the quality of their threat models. For instance, Participant \textbf{P2} uses peer-evaluation to assess the quality of the threat models that they create. They Said \textit{" There were certain folks that we would do peer reviews [of their] threat models."}. Participant \textbf{P3} performs review at each project milestone to ensure the work done at the given milestone is of sufficient quality. They said \textit{"at each of the gates or milestones, you do the proper review to make sure that the work that was done up until that point is of sufficient quality."} And, Participant \textbf{P6} uses a set of requirements to verify the coverage of the developed threat model of the important security aspects related to the domain of the given system.
 
\subsection{Roles Involved in Threat Modeling}

\begin{table}
\centering
\caption{Roles in the threat modeling processes.}\label{tab:people}
\rowcolors{2}{gray!25}{white}
\renewcommand{\cellalign}{l}
\small
\begin{tabular}{p{0.6in} p{2.25in}}
\hline 
\rowcolor{gray!50}
Role & Description \\
\hline 
Security team & Initiate the threat modeling process and perform the threat modeling exercise.\\

Architect & Provide the documentation and artifacts about the system. The security team may interview them to get more details about the system.\\

Developer & The security team interviews the developers to get more details about the system.\\

Stakeholder &  The security team interviews the other stakeholders of a system as needed to get more details.\\

\hline 
\end{tabular}
\vspace{-2em}
\end{table}

Table \ref{tab:people} lists the common roles that the participants work with when performing threat modeling. Some of the participants involve the \ac{CPS} operators, the management staff, the subject matter experts, and the equipment suppliers in their threat modeling exercise as they need. These roles help to gain depth understanding of the system, including the different environments of running the given \ac{CPS}, the operations of the system, the used equipment, and possibly other aspects. Interviewing different stockholders helps to develop a "complete" threat model.

\subsection{Tool}

Three participants use Microsoft Threat Modeling Tool~\cite{tmt} although the tool does not cover the physical components of \acp{CPS} and three participants use their own tools, including custom templates, for threat modeling. For instance, {\bf P9} said \textit{"Microsoft has a threat modeling tool [...], and there is actually an automotive template that we look at to plug into our system."}

\begin{figure*}
    \centering
    \fbox{
    \begin{minipage}{14 cm}
\tikzset{every picture/.style={line width=0.75pt}} 

\begin{tikzpicture}[x=0.75pt,y=0.75pt,yscale=-1,xscale=1]

\draw    (354.3,290.3) -- (422.26,290.3) ;
\draw [shift={(425.26,290.3)}, rotate = 180] [fill={rgb, 255:red, 0; green, 0; blue, 0 }  ][line width=0.08]  [draw opacity=0] (8.93,-4.29) -- (0,0) -- (8.93,4.29) -- cycle    ;
\draw    (180,220.2) -- (232.13,272.86) -- (236.61,277.39) ;
\draw  [color={rgb, 255:red, 183; green, 212; blue, 247 }  ,draw opacity=1 ][fill={rgb, 255:red, 183; green, 212; blue, 247 }  ,fill opacity=1 ] (234.3,204.52) .. controls (234.3,199.81) and (238.11,196) .. (242.82,196) -- (355.78,196) .. controls (360.49,196) and (364.3,199.81) .. (364.3,204.52) -- (364.3,230.08) .. controls (364.3,234.79) and (360.49,238.6) .. (355.78,238.6) -- (242.82,238.6) .. controls (238.11,238.6) and (234.3,234.79) .. (234.3,230.08) -- cycle ;
\draw  [color={rgb, 255:red, 183; green, 212; blue, 247 }  ,draw opacity=1 ][fill={rgb, 255:red, 183; green, 212; blue, 247 }  ,fill opacity=1 ] (235.3,279.52) .. controls (235.3,274.81) and (239.11,271) .. (243.82,271) -- (356.78,271) .. controls (361.49,271) and (365.3,274.81) .. (365.3,279.52) -- (365.3,305.08) .. controls (365.3,309.79) and (361.49,313.6) .. (356.78,313.6) -- (243.82,313.6) .. controls (239.11,313.6) and (235.3,309.79) .. (235.3,305.08) -- cycle ;
\draw  [color={rgb, 255:red, 183; green, 212; blue, 247 }  ,draw opacity=1 ][fill={rgb, 255:red, 183; green, 212; blue, 247 }  ,fill opacity=1 ] (426.26,281.52) .. controls (426.26,276.81) and (430.07,273) .. (434.78,273) -- (547.74,273) .. controls (552.45,273) and (556.26,276.81) .. (556.26,281.52) -- (556.26,307.08) .. controls (556.26,311.79) and (552.45,315.6) .. (547.74,315.6) -- (434.78,315.6) .. controls (430.07,315.6) and (426.26,311.79) .. (426.26,307.08) -- cycle ;
\draw  [color={rgb, 255:red, 183; green, 212; blue, 247 }  ,draw opacity=1 ][fill={rgb, 255:red, 183; green, 212; blue, 247 }  ,fill opacity=1 ] (40.3,229.52) .. controls (40.3,224.81) and (44.11,221) .. (48.82,221) -- (161.78,221) .. controls (166.49,221) and (170.3,224.81) .. (170.3,229.52) -- (170.3,255.08) .. controls (170.3,259.79) and (166.49,263.6) .. (161.78,263.6) -- (48.82,263.6) .. controls (44.11,263.6) and (40.3,259.79) .. (40.3,255.08) -- cycle ;
\draw    (181.3,241.3) -- (172.5,241.37) ;
\draw [shift={(169.5,241.4)}, rotate = 359.51] [fill={rgb, 255:red, 0; green, 0; blue, 0 }  ][line width=0.08]  [draw opacity=0] (8.93,-4.29) -- (0,0) -- (8.93,4.29) -- cycle    ;
\draw    (232.61,309.39) -- (170.95,342.38) ;
\draw [shift={(168.3,343.8)}, rotate = 331.85] [fill={rgb, 255:red, 0; green, 0; blue, 0 }  ][line width=0.08]  [draw opacity=0] (8.93,-4.29) -- (0,0) -- (8.93,4.29) -- cycle    ;
\draw  [color={rgb, 255:red, 183; green, 212; blue, 247 }  ,draw opacity=1 ][fill={rgb, 255:red, 183; green, 212; blue, 247 }  ,fill opacity=1 ] (426.3,351.3) .. controls (426.3,346.59) and (430.11,342.78) .. (434.82,342.78) -- (547.78,342.78) .. controls (552.49,342.78) and (556.3,346.59) .. (556.3,351.3) -- (556.3,376.86) .. controls (556.3,381.57) and (552.49,385.38) .. (547.78,385.38) -- (434.82,385.38) .. controls (430.11,385.38) and (426.3,381.57) .. (426.3,376.86) -- cycle ;
\draw  [color={rgb, 255:red, 183; green, 212; blue, 247 }  ,draw opacity=1 ][fill={rgb, 255:red, 183; green, 212; blue, 247 }  ,fill opacity=1 ] (234.3,350.52) .. controls (234.3,345.81) and (238.11,342) .. (242.82,342) -- (355.78,342) .. controls (360.49,342) and (364.3,345.81) .. (364.3,350.52) -- (364.3,376.08) .. controls (364.3,380.79) and (360.49,384.6) .. (355.78,384.6) -- (242.82,384.6) .. controls (238.11,384.6) and (234.3,380.79) .. (234.3,376.08) -- cycle ;
\draw    (361.3,313.3) -- (425.13,344.03) ;
\draw [shift={(427.83,345.33)}, rotate = 205.71] [fill={rgb, 255:red, 0; green, 0; blue, 0 }  ][line width=0.08]  [draw opacity=0] (8.93,-4.29) -- (0,0) -- (8.93,4.29) -- cycle    ;
\draw  [color={rgb, 255:red, 183; green, 212; blue, 247 }  ,draw opacity=1 ][fill={rgb, 255:red, 183; green, 212; blue, 247 }  ,fill opacity=1 ] (40.3,179.12) .. controls (40.3,174.41) and (44.11,170.6) .. (48.82,170.6) -- (161.78,170.6) .. controls (166.49,170.6) and (170.3,174.41) .. (170.3,179.12) -- (170.3,204.68) .. controls (170.3,209.39) and (166.49,213.2) .. (161.78,213.2) -- (48.82,213.2) .. controls (44.11,213.2) and (40.3,209.39) .. (40.3,204.68) -- cycle ;
\draw    (299.3,240.3) -- (299.3,266.3) ;
\draw [shift={(299.3,269.3)}, rotate = 270] [fill={rgb, 255:red, 0; green, 0; blue, 0 }  ][line width=0.08]  [draw opacity=0] (8.93,-4.29) -- (0,0) -- (8.93,4.29) -- cycle    ;
\draw    (299.3,316.3) -- (299.3,337.3) ;
\draw [shift={(299.3,340.3)}, rotate = 270] [fill={rgb, 255:red, 0; green, 0; blue, 0 }  ][line width=0.08]  [draw opacity=0] (8.93,-4.29) -- (0,0) -- (8.93,4.29) -- cycle    ;
\draw    (181,190) -- (180.5,241.4) ;
\draw  [color={rgb, 255:red, 183; green, 212; blue, 247 }  ,draw opacity=1 ][fill={rgb, 255:red, 183; green, 212; blue, 247 }  ,fill opacity=1 ] (40.3,348.52) .. controls (40.3,343.81) and (44.11,340) .. (48.82,340) -- (161.78,340) .. controls (166.49,340) and (170.3,343.81) .. (170.3,348.52) -- (170.3,374.08) .. controls (170.3,378.79) and (166.49,382.6) .. (161.78,382.6) -- (48.82,382.6) .. controls (44.11,382.6) and (40.3,378.79) .. (40.3,374.08) -- cycle ;
\draw    (180.5,219.4) -- (233.61,219.39) ;
\draw    (181.3,190.3) -- (172.5,190.37) ;
\draw [shift={(169.5,190.4)}, rotate = 359.51] [fill={rgb, 255:red, 0; green, 0; blue, 0 }  ][line width=0.08]  [draw opacity=0] (8.93,-4.29) -- (0,0) -- (8.93,4.29) -- cycle    ;
\draw    (489.3,317.3) -- (489.3,338.3) ;
\draw [shift={(489.3,341.3)}, rotate = 270] [fill={rgb, 255:red, 0; green, 0; blue, 0 }  ][line width=0.08]  [draw opacity=0] (8.93,-4.29) -- (0,0) -- (8.93,4.29) -- cycle    ;
\draw    (181.3,219.4) -- (232.3,180.2) -- (489.3,181.2) -- (490.3,271.2) ;

\draw (301.3,218.3) node  [font=\footnotesize] [align=left] {Cyber-physical system};
\draw (392,281.6) node  [font=\footnotesize] [align=left] {include};
\draw (492.26,292.3) node  [font=\footnotesize] [align=left] { \ \ Continuous threat \\ \ \ \ \ \ modeling};
\draw (106.3,360.3) node  [font=\footnotesize] [align=left] {Roles};
\draw (333,253.6) node  [font=\footnotesize] [align=left] {object of};
\draw (303.3,292.3) node  [font=\footnotesize] [align=left] {\\Threat model};
\draw (104.3,243.3) node  [font=\footnotesize] [align=left] {Threat business impacts};
\draw (491.3,362.08) node  [font=\footnotesize] [align=left] {Tools };
\draw (405,320.6) node  [font=\footnotesize] [align=left] {use};
\draw (104.3,191.9) node  [font=\footnotesize] [align=left] {Security properties};
\draw (213,211.6) node   [align=left] {{\footnotesize require}};
\draw (505,327.6) node  [font=\footnotesize] [align=left] {use};
\draw (313,325.6) node  [font=\footnotesize] [align=left] {use};
\draw (299.3,363.3) node  [font=\footnotesize] [align=left] { \\ \ \ \ \ Threat modeling \\ approaches and methods};
\draw (235,253.6) node   [align=left] {{\footnotesize verify}};
\draw (190,314.6) node   [align=left] {{\footnotesize involve}};
\draw (353,166.6) node   [align=left] {{\footnotesize verify}};

\end{tikzpicture}

 \end{minipage}
}
\caption{Model of the threat modeling concepts and the relationships among these concepts.} 

\label{fig:model}
\vspace{-1em}
\end{figure*}
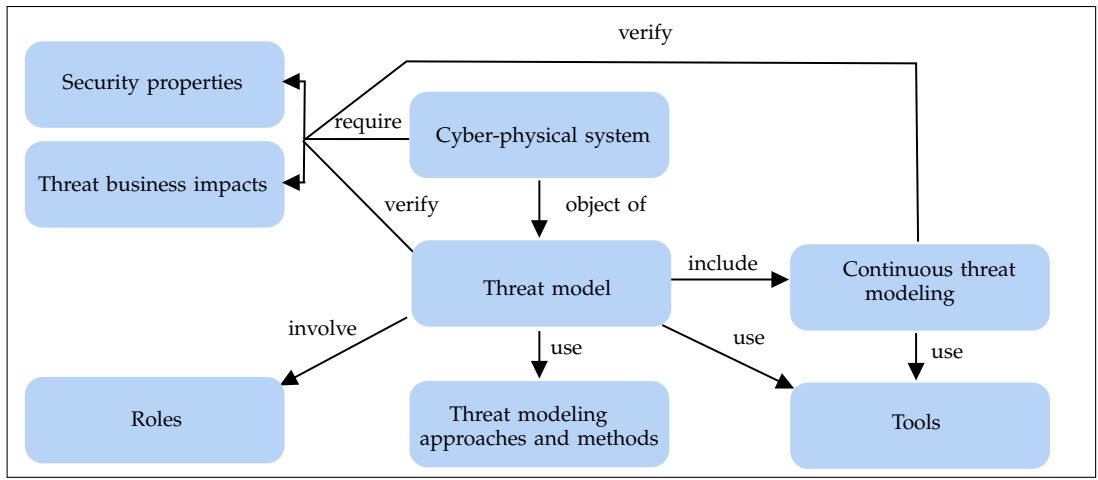

\subsection{Challenges}

The participants reported few challenges that they face when performing threat modeling of \acp{CPS}, which we discuss in the following.

\noindent{\bf Variety of \acp{CPS}.}
Several of the participants had to work on threat models of \acp{CPS} for several applications domains (e.g., mining, transportation, smart grid) and use a variety of physical components that are often not familiar with at the beginning of the projects. They find it impossible to have broad knowledge about threats for \acp{CPS} and difficult to generalize expertise across \acp{CPS}'s application domains.\footnote{This different from IT systems that use known architecture styles and follow standard components definitions, e.g., web applications.} Participants that have IT background find themselves with limited knowledge about the physical components: they are not familiar with the threats to the system that they analyze and to the mechanisms that could be utilized to mitigate the threats to these systems. Some participants proposed developing a repository of patterns and mitigation strategies since there are many threat vectors and attack agents to consider.

\noindent  {\bf Limitation of current threat modeling approaches and methods.} The existing threat modeling approaches, such as STRIDE and PASTA, focus on computer security. The use of these methods to perform threat modeling for \acp{CPS} may produce incomplete threat models because these methods do not cover the physical aspects of \acp{CPS}. Some participants suggest the development of a framework that allows identifying common practical attack scenarios based on the application domains of \acp{CPS}.

\noindent {\bf Limitation of tools.} Microsoft Threat Modeling Tool is commonly used to generate an initial list of threats to a given system based on a default template that uses the STRIDE taxonomy. It is known that STRIDE focuses on computer security threats; hence it would produce incomplete threat models for \acp{CPS}.

\noindent {\bf Challenge in current culture.} Current business culture of "publish now and fix later" has been a challenge for some participants--sometimes only the threats that are related to publicly known attacks are considered. To address this problem, Participant \textbf{P4} proposes to have the security experts develop quality threat models that use publicly available threat patterns. They said: \textit{"I think that would be very useful for the industry at large is a set of threat model patterns."}

\section{Discussion}\label{sec:Discussion}
This section summarizes the results of the study and discusses the impacts of the study and its limitations.

\subsection{Summary}

Figure~\ref{fig:model} shows the themes extracted from the study and the relationships among these themes. The figure shows that \acp{CPS} have security properties requirements and other associated requirements such as safety. The goal of the threat modeling processes and the continuous threat modeling sub-processes is to identify and rank system weaknesses that violate these requirements. The participants use several threat modeling methods and approaches and involve several stakeholders of the \acp{CPS} that they perform threat models of.

According to the participants in our study, integrity and availability are the security properties the most utmost of concern for \acp{CPS}. In addition, many participants use the threat modeling method STRIDE, which is unexpected since the method focuses on the threats to IT systems, not \acp{CPS}. Also, most of the participants use a combination of known approaches, known methods, and known standards when performing threat modeling of a \ac{CPS}. We note that the participants associate the quality of threat models mainly to the skills and experience of the security experts who perform the threat modeling. The two techniques that some participants use to ensure the quality of threat models developed by their subordinates are peer-evaluation and the use of the quality checklist.   

\subsection{Impact of the Study}

Existing threat taxonomies, such as STRIDE, focus on either the CIA triad or the controllability of the physical components of a system. This study reveals that experts focus on the threats to the integrity, availability, controllability, and safety of the systems when performing threat modeling of \acp{CPS}. The community should develop a knowledge repository of practical threats to \acp{CPS} that consider the business impacts of failure of physical components, including safety besides the CIA triad.

We found that most of the participants use a combination of known threat modeling approaches, methods, and standards, which makes threat modeling time consuming--it is done two or more times. This calls for developing \emph{practical} new threat modeling approaches that integrate both the IT and OT security needs of \acp{CPS} effectively. Such methods should help security practitioners to produce quality threat models for \ac{CPS} that could be trusted by the project managers.

We also observed that the participants use their own template to tie the risk to the threats of \acp{CPS}. Developing risk assessment methods for \acp{CPS} acceptable by the major actors in the industry will help the experts to communicate better and exchange information about risks of \acp{CPS}.

In addition, we found that most of the participants do not use quality assurance methods for the threat models that they produce. The managers sometimes request threat models for their \acp{CPS} from more than one experts, especially when the system gets hacked. The community should explore techniques and standards for assessing the quality of threat models.

\subsection{Threats to Validity}

Initially, we gave an open-source of a \ac{CPS} to some of the selected participants and hoped that they provide us with their threat models, which we could use to study the practice of threat modeling in depth. The volunteer participants did not want that given, among others, the required high time commitment to do so. Therefore, we opted for exploitative interviews to address our research questions. 

The limitations of the study are classified into construct validity, internal validity, conclusion validity, and external validity are discussed as the following ~\cite{CrOt2017,WRHO2000}.

\noindent{\bf Construct validity.} To address the validity of the relations between the performed study and the goal of the study, we performed a literature review, designed an interview protocol, and tested it with some experts. We collected information from eleven participants who have different roles and are located in different cities. This gives confidence in the stability of the collected data.

\noindent{\bf Internal validity.} To address the validity of the relationship between the study and its results, we tell the participants at the beginning of the interviews the goals of the interview, which should help in ensuring that the participant and the interviewer share the same goal. 

\noindent{\bf Conclusion validity.} To address the validity of the ability to make correct conclusions from the results of the study, the main author provided the second author their codes and the themes for each of the interview, who reviewed them, to reduce the subjectivity of the results.

\noindent{\bf External validity.} To address the validity of the generalization of the study, the eleven participants in the study are selected to be security experts from nine organizations in different businesses. We believe the diverse experience of the participants supports generalizing the results.

\section{Conclusion}\label{sec:Conclusion}
This paper reports about the practice of threat modeling of \acp{CPS}. We conclude that (1) ensuring the integrity and availability of data and system's components in addition to controllability and safety of \acp{CPS} is the concern of threat modeling of \acp{CPS}, (2) there are differences between experts with a background in control system and experts with a background in IT regarding the approaches to perform threat modeling, (3) the experts use a combination of known approaches, methods, and standards to perform threat modeling of a given \ac{CPS}, (4) most of the threat modeling participants perform continuous threat modeling, (5) the experts often use custom risk scoring methods, (6) most of the participants do not use quality assurance techniques for the threat models that they produce and rely depend on the experience and skills of the expert who performs the threat model, and (7) four roles are commonly involved in threat modeling, namely security team, architect, developer, and stakeholder.

\section*{Acknowledgment}
The author thank Simone Curzi from Microsoft,  Zafar Ali from John Deere, Rohini Narasipur from Bosch, Arun Prabhakar from Security Compass, and Youssef Jad from PM SCADA Cyber Defense for participating in the study and reviewing this paper. The authors thank also the other anonymous participants in the study for their contributions. The participants were not representing their respective employers in the study. 
\bibliographystyle{ieeetr}
\bibliography{Sections/ThreatModel}

\end{document}